\documentclass[aps, prd, twocolumn, nofootinbib, longbibliography]{revtex4-1}

\usepackage{epsfig}
\usepackage{slashed, enumitem}
\usepackage[utf8]{inputenc}

\usepackage{float,color}
\usepackage{subfigure}
\usepackage{multirow}
\usepackage{slashed}
\usepackage{placeins} 
\usepackage{mathrsfs} 

\usepackage{amsmath}
\usepackage{amssymb}
\usepackage{graphicx}

\usepackage{slashed}
\usepackage{multirow}
\usepackage[dvipsnames]{xcolor}
\usepackage{epstopdf}
\usepackage{soul}

\usepackage{siunitx}
\usepackage{xspace}

\usepackage{subfigure}

\usepackage{upgreek}
\usepackage{mathtools}
\usepackage{setspace, slashed, gensymb, hhline, booktabs}
\usepackage{wrapfig}
\usepackage{makecell}

\usepackage{mhchem}

\definecolor{lime}{HTML}{A6CE39}
\definecolor{cc}{HTML}{F72585}

\usepackage[colorlinks=true, citecolor=blue, linkcolor=blue]{hyperref}	

\usepackage{tikz,xcolor}

\definecolor{lime}{HTML}{A6CE39}
\DeclareRobustCommand{\orcidicon}{\hspace{-4pt}
	\begin{tikzpicture}
		\draw[lime, fill=lime] (0,0) 
		circle [radius=0.16] 
		node[white] {\hspace{0.1mm}{\fontfamily{qag}\selectfont \tiny ID}};
		\draw[white, fill=white] (-0.07,0.1) 
		circle [radius=0.01];
	\end{tikzpicture}
	\hspace{-3.2mm}
}

\foreach \x in {A, ..., Z}{\expandafter\xdef\csname 
	orcid\x\endcsname{\noexpand\href{https://orcid.org/\csname orcidauthor\x\endcsname}
		{\noexpand\orcidicon}}
}

\begin{document}

\title{Using supernova neutrinos to probe strange spin of proton \\ with JUNO and THEIA}
 
\author{Bhavesh Chauhan\orcidA{} }
\email{bhavesh-chauhan@uiowa.edu}
\affiliation{Department of Physics and Astronomy, University of Iowa, Iowa City, IA 52242, USA}

\date{\today}

\begin{abstract}
The strange quark contribution to proton's spin ($\Delta s$) is a fundamental quantity that is poorly 
determined from current experiments. Neutrino-proton elastic scattering (pES) is a promising channel 
to measure this quantity, and requires an intense source of low-energy neutrinos and a low-threshold 
detector with excellent resolution. In this paper, we propose that neutrinos from a galactic 
supernova and their interactions with protons in large-volume scintillation detectors can be utilized to 
determine $\Delta s$. The spectra of all flavors of supernova neutrinos can be independently 
determined using a combination of DUNE and Super-(Hyper-)Kamiokande. This 
allows us to predict pES event rates in JUNO and THEIA, and estimate $\Delta s$ by comparing with detected 
events. We find that the projected sensitivity for a supernova at 1 kpc (10 kpc), is approximately $\pm 0.01$ 
($\pm 0.15$). Interestingly, the limits from a nearby supernova would be comparable to the 
results from lattice QCD, and better than polarized deep-inelastic scattering experiments. Using supernova 
neutrinos provides a true $Q^2\rightarrow 0$ measurement, and thus an axial-mass independent 
determination of $\Delta s$. 
\end{abstract}

\maketitle

\section{Introduction}
One of the very first measurement of the contribution of strange quarks to the spin of proton was 
performed by the European Muon Collaboration (EMC). The strange spin 
of proton ($\Delta s$) is related to 
the axial form-factor in the limit of vanishing momentum-transfer ($Q^2$). Usually, the experiments measure 
$\Delta s$ at $Q^2 \sim \rm GeV^2 $, and extrapolate the results to $Q^2=0$ by assuming some 
parametrization for the 
axial form-factor. The EMC collaboration obtained, 
\begin{equation}
	\nonumber
	\Delta s \,{\rm(EMC)}=  - 0.095 \pm 0.016 \pm 0.023
\end{equation}
through polarized deep-inelastic scattering (pDIS) \cite{EuropeanMuon:1989yki}. The contribution is 
not only non-zero, which is in violation of the Ellis-Jaffe 
sum rule \cite{Ellis:1973kp}, but also negative. Subsequent measurements by COMPASS 
\cite{COMPASS:2006mhr} and HERMES \cite{HERMES:2006jyl} have independently determined 
$\Delta s$, and their results are consistent with EMC. A recent global analysis can be found in Ref. 
\cite{Leader:2014uua}. 

On the theory front, heavy-quark contribution to the axial form-factor has been estimated in Refs. 
\cite{Collins:1978wz, Mohapatra:1978bc, Wolfenstein:1978wd}. In recent times, precise lattice QCD 
calculations of the nuclear structure suggest that  
\begin{equation}
	\nonumber
	\Delta s \,{\rm(Lattice)}=  - 0.018 \pm 0.006
\end{equation}
which is closer to zero, but negative \cite{Chambers:2015bka}. Other groups have also obtained similar 
estimates (see Refs. \cite{Dong:1995rx, QCDSF:2011aa, Engelhardt:2012gd}).

Since neutrinos only interact via the weak force, neutrino proton elastic scattering (pES) is sensitive to 
the axial form-factor and a possible channel for measuring $\Delta s$. A clear advantage of using 
neutrinos is that one does not need to rely on flavor symmetries or fragmentation functions. With this 
motivation, the E734 experiment at Brookhaven National Lab used neutrino beams on a liquid 
scintillator target to determine the axial form-factor for $Q^2\in0.45$--1.05 
$\rm{GeV}^2$ \cite{Ahrens:1986xe}. Upon 
extrapolating their results to $Q^2=0$, they obtained 
\begin{equation}
	\nonumber
	\Delta s \,{\rm(E734)}=  - 0.15 \pm 0.09
\end{equation}
for fixed axial-mass parameter ($M_A$). If $M_A$ is not fixed, then there is large 
uncertainty in the extracted value of $\Delta s$ as they are strongly correlated at these energies. The 
Fermilab experiment MiniBooNE has also measured  $\Delta s$ through 
pES and obtained $\Delta s= 0.08 \pm 0.26 $ by measuring the ratio of neutrino-proton and 
neutrino-nucleon interactions. In Ref. \cite{Sufian:2018qtw}, the authors 
reanalyze MiniBooNE data using $z$-expansion parametrization of the form factors and obtain 
\begin{equation}
	\nonumber
	\Delta s \,{\rm(MiniBooNE)}=  - 0.102 \pm 0.178 \pm 0.080
\end{equation}
where the uncertainty is smaller due to fixing some parameters with results from lattice QCD. For a 
measurement of $\Delta s$ that is practically independent of $M_A$, one requires an intense source 
of low-energy neutrinos and a detector that can efficiently measure the small energies of the recoiling 
proton. In Ref. \cite{Pagliaroli:2012hq}, the authors propose using neutrinos from pion deay-at-rest to 
measure $\Delta s$ in a kton-scale scintillation detector. In this paper, we look at the possibility of 
measuring $\Delta s$ using supernova neutrinos and large-volume scintillation detectors. 

In a core collapse supernova, nearly 99\% of the binding energy of the progenitor is released in 
the form of neutrinos. These neutrinos are emitted with energies $\mathcal{O}(10)$ MeV, and can 
generate nuclear recoils with $Q^2 \sim 0.01\,{\rm GeV}^2$. The elastic scattering of these neutrinos with 
protons has been identified as a promising channel to study 
the non-electron flavors, i.e., $\nu_x = \nu_\mu,\,\bar{\nu}_\mu,\,\nu_\tau,\,\bar{\nu}_\tau$, using 
scintillation detectors \cite{Beacom:2002hs}. Even though the scintillation from recoil proton is 
quenched, one can get enough statistics to reconstruct the spectrum of 
$\nu_x$ \cite{Dasgupta:2011wg,Lu:2016ipr, Li:2017dbg, Li:2019qxi, Nagakura:2020bbw}. In these 
reconstruction techniques, $\Delta s$ has been regarded as a source of uncertainty 
in the cross-section. In this paper, we invert the question and ask, \emph{``how well can we constrain 
$\Delta s$ using neutrinos from a galactic supernova?"}.  The 
uncertainty in reconstruction of $\Delta s$ can be approximated by, 
\begin{equation}
	\delta{\Delta s} \approx \frac{1.27}{2} \left[ \frac{1}{N}  + \left( 
	\frac{\delta \Phi}{\Phi}  \right)^2     \right]^{1/2}
\end{equation} 
where $N$ is the total number of events and $\delta \Phi/\Phi$ is the fractional uncertainty in the \emph{total} 
supernova neutrino flux. A multi-kton scintillation detector would observe few thousand pES events 
from a supernova at 10 kpc, resulting in a small contribution from the first term. Hence, the 
reconstruction of $\Delta s$ mainly depends on how well we can determine the supernova neutrino 
spectra. 

Of all currently operational neutrino telescopes, the most promising candidate to measure supernova 
neutrino spectra is the water Cherenkov detector Super-Kamiokande \cite{Ikeda:2007sa}. The addition 
of Gadolinium (Gd) salts has greatly enhanced the neutron tagging efficiency \cite{Beacom:2003nk, 
Super-Kamiokande:2021the} and allows for a clean and large-statistics measurement of $\bar{\nu}_e$ 
using inverse beta decay. In order to detect other flavors, one has to rely on future detectors such as 
DUNE \cite{Abi:2020wmh, Abi:2020evt}, JUNO \cite{An:2015jdp,  Djurcic:2015vqa, 
Abusleme:2021zrw}, Hyper-Kamiokande \cite{Abe:2011ts}, and THEIA 
\cite{Askins:2019oqj}. There are 
other proposed detectors such as HALO \cite{Duba:2008zz}, deuterated liquid scintillator 
\cite{Chauhan:2021snf}, and RES-NOVA \cite{RES-NOVA:2021gqp} that have relatively smaller volume, 
but will play an important role in detection of supernova neutrinos. 

For our analysis, we find that DUNE is the most promising candidate for measurement of $\nu_e$ 
spectrum via charged-current interactions with argon \cite{Nikrant:2017nya}. The water Cherenkov 
detector Super-Kamiokande (SK) or its upgrade Hyper-Kamiokande (HK) (both assumed to have Gd) 
would measure $\bar{\nu}_e$ spectrum via inverse-beta decay. The non-electron flavors, i.e., $\nu_x$ 
can only be detected through neutral-current interactions. In SK and HK, $\nu_x$ will be detected 
using elastic scattering with electrons, which also gets contribution from $\nu_e$ and $\bar{\nu}_e$ 
\cite{GalloRosso:2017mdz}. Thus, a combination of DUNE and SK/HK, can detect all three components 
of supernova neutrinos independent of $\Delta s$. The measured neutrino spectra 
from these detectors can be used to predict the pES event rates in scintillation detectors, and 
comparison with measured rates can yield limits on $\Delta s$. This possibility is explored in detail in 
this paper.  

\begin{figure}[t]
	\includegraphics[width=0.42\textwidth]{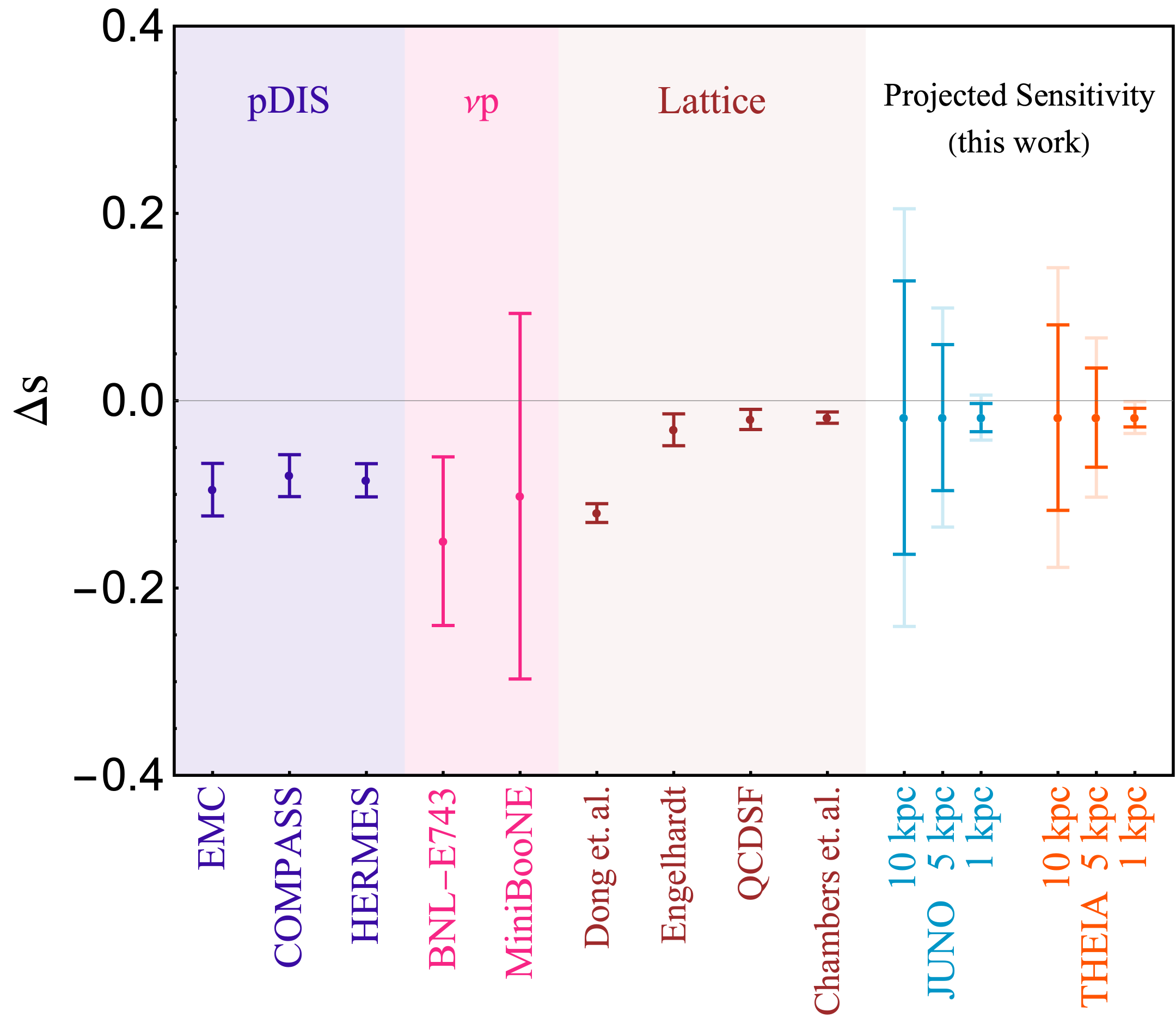}
	\caption{\label{fig:main} The $1\sigma$ estimate of $\Delta s$ from the polarized deep inelastic 
		scattering (pDIS) experiments, $\nu$p scattering experiments, as well as Lattice QCD are shown 
		along with our projected sensitivity from a supernova at 10, 5, and 1 kpc. The darker blue and 
		orange limits are for JUNO and THEIA respectively where the fluence parameters are determined 
		from DUNE+HK. The weaker limits obtained using reconstructed fluence parameters DUNE+SK are 
		shown with a lighter shade. }
\end{figure}

So far, only linear alkyl benzene (LAB) based scintillator detectors such as Borexino, KamLAND, SNO+, 
and JUNO have been considered as candidates to detect pES events from a galactic supernova 
\cite{Beacom:2002hs, Dasgupta:2011wg, Lu:2016ipr, Li:2017dbg, Li:2019qxi}. In this paper, we show 
that the water-based liquid scintillator (WbLS) detector, THEIA, would be an excellent candidate to 
detect pES events. The higher light yield and smaller $\,^{14}\rm{C}$ concentration allows for a lower 
threshold as compared to JUNO, and larger events rates despite similar volume. Using a benchmark 
value for fluence parameters, and uncertainties from DUNE and SK/HK determined from Ref. 
\cite{Nikrant:2017nya} and \cite{GalloRosso:2017mdz}, we determine the sensitivity to $\Delta s$ using 
a simple Monte Carlo simulation for uncertainty quantification. The main results of this paper are summarized 
in 
Fig. \ref{fig:main}, where we show our projected sensitivity from a galactic supernova along with results from 
pDIS, beam-neutrino experiments, and lattice QCD.  

\section{Forecast for ${\rm pES}$ in THEIA}\label{sec:forecast}

THEIA is a proposed hybrid detector that uses a cocktail-like water-based liquid scintillator (WbLS) as 
target \cite{Askins:2019oqj}. There are two possible configurations, THEIA25 with 25 kton WbLS (20 
kton fiducial volume) and THEIA100 with 100 kton WbLS (80 kton fiducial volume). In this work, we 
provide estimate for THEIA25, with an understanding that the results for THEIA100 can be 
appropriately scaled. The unique advantage of WbLS is the ability to simultaneously detect both 
scintillation and Cherenkov signals of charged particles. A large-scale WbLS detector will have 
excellent capacity to study diffuse supernova neutrino background, reactor neutrinos, and new physics 
scenarios such as proton decay \cite{Askins:2019oqj}. If operational at the time, THEIA can also detect 
neutrinos from the 
next galactic supernova through inverse beta decay, electron elastic scattering, and interactions with 
oxygen (similar to a Cherenkov detector like SK/HK). As THEIA can also operate as 
scintillation detector, it will be sensitive to pES similar to JUNO.

To estimate the spectrum of pES events in THEIA, we follow the outline of Ref. \cite{Dasgupta:2011wg}. We 
assume a fiducial galactic supernova that emits a total energy of $3 \times 10^{53}$\,erg over a duration of 
$\sim15\,$s. The 
fluence of neutrinos from such a supernova is given by
\begin{equation}
\label{eq:fluence}
\dfrac{dF_\nu}{dE} = \frac{1}{4 \pi d^2} \frac{\mathcal{E_\nu}}{\langle E_\nu \rangle} \times 
\frac{d\varphi_\nu}{dE}
\end{equation} 
where $\nu = \nu_e, \bar{\nu}_e, \nu_x$ represents the neutrino flavors, $d$ is the distance to the supernova, 
$\mathcal{E}_\nu$ is the luminosity, $\langle E_\nu \rangle$ is the average energy, and 
$d\varphi_\nu/dE$ is the unit-normalized spectra which can be parameterized as, 
\begin{equation}
 \dfrac{d\varphi_\nu}{dE} = \frac{(1 + \alpha_\nu)^{1 + 
			\alpha_\nu}}{\Gamma(1 + \alpha_\nu)} \frac{E^{\alpha_\nu}}{\langle E_\nu \rangle^{1 + 
			\alpha_\nu}} 
			\exp\left( - (1 + \alpha_\nu) \frac{E}{\langle E_\nu \rangle} \right)
\end{equation}
where $\alpha_\nu $ is called the \emph{pinching} parameter \cite{Keil:2002in}. For our estimates, 
we use $d=10$ kpc, $\alpha_\nu$ = 3, $\langle E_{\nu_e} \rangle$=12 MeV, $\langle E_{\bar{\nu}_e} 
\rangle$=14 MeV, and $\langle E_{\nu_x} \rangle$=16 MeV. We assume equipartition of among all six 
flavors.

The differential cross section for pES in the Llewellyn-Smith formalism is
\begin{equation}\label{diffxs}
	\frac{d\sigma}{dQ^2} = \frac{G_F^2 M_p^2}{8 \pi E_\nu^2} \left[	A \pm B \frac{s-u}{M_p^2} + C 
	\frac{(s-u)^2}{M_p^4}\right]\,,
\end{equation}
where $s-u = 4 M_p E_\nu - Q^2$, $Q^2 = 2 M_p T_p$, and $T_p$ is the kinetic energy of the recoiling 
proton \cite{LlewellynSmith:1971uhs}. In terms of $\tau = Q^2/4M_p^2$, 
\begin{align}
	\nonumber 
	A &= 4 \tau\left[(1+\tau)G_A^2-(1-\tau)F_2^2+\tau(1-\tau)F_2^2+4 \tau F_1 F_2\right],\\ 
	\nonumber 
	B& = 4 \tau \left[ G_A (F_1 + F_2)\right],\\
	\nonumber
	C&= \frac14 \left[ G_A^2 + F_1^2 + F_2^2 \right],
\end{align}
where $F_{1,2}$ are the vector form-factors\footnote{We use $\sin^2(\theta_W)$ = 0.238, which is 
the appropriate low-$Q^2$ limit, resulting in a slightly weaker dependence on vector form-factors as 
compared to Ref. \cite{Beacom:2002hs}.}, and $G_A$ is the 
axial form-factor \cite{Ahrens:1986xe}. For supernova neutrino interactions, the maximum kinetic 
energy of 
recoiling protons is only a few MeV and a simpler expression for the cross section can be derived in the limit 
$Q^2 \rightarrow 0$ \cite{Beacom:2002hs}. The axial form-factor in this limit can be written as, 
\begin{equation}
	\lim_{Q^2 \rightarrow 0} G_A  = \frac12 \left( \mathtt{g}_A - \Delta s\right)
\end{equation}
where $\mathtt{g}_A = 1.2755(11)$ \cite{Czarnecki:2018okw}. We note that, for $\Delta s \in \pm0.15$, 
the total cross section can change by $\mp\,25\%$. To estimate the event rates, we use 
$\Delta s = - 0.018$, which is the central value reported in Ref. \cite{Chambers:2015bka}.

\begin{figure} [t]
	\includegraphics[width=0.4\textwidth]{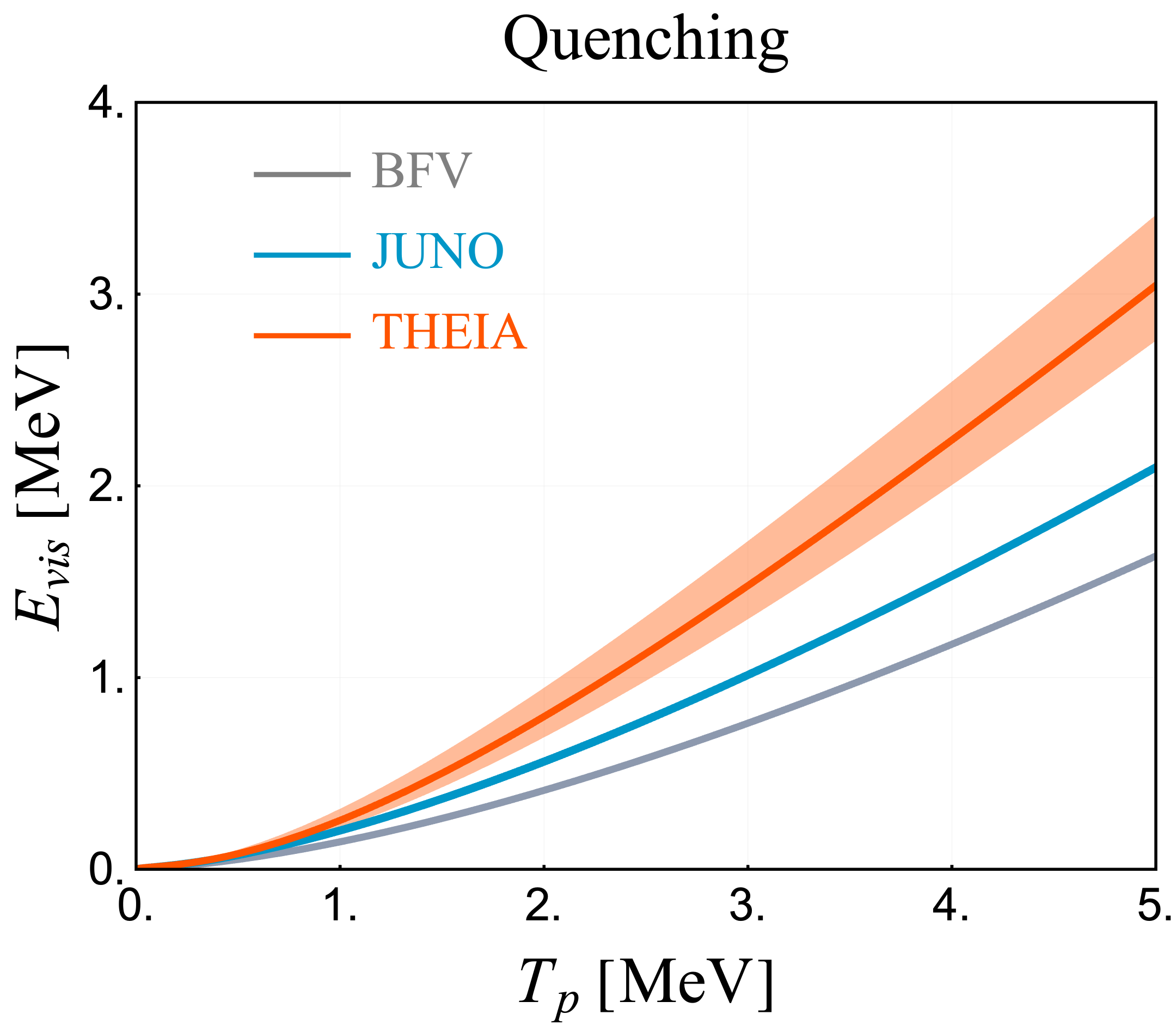}
	\caption{\label{fig:quenching} The visible energy ($E_{vis}$) as a function of the recoil proton 
		energy in JUNO (from von Krosigk et. al. \cite{vonKrosigk:2013sa}) and in THEIA 
		(from Callagan et. al. \cite{Callaghan:2022ahi}) is shown in blue and orange respectively. The shaded 
		region shows the $1\sigma$ uncertainty arising from $k_B$ and $C$ given in eq. \eqref{eq:unc}. 
		For comparison, we also show in gray the quenching for $k_B$ 
		= 0.015 cm/MeV which is used in Ref. \cite{Beacom:2002hs}.}
\end{figure}

Only a fraction of proton's recoil energy is converted into detectable scintillation signal, a phenomenon known 
as \emph{quenching}. The quenched proton energy, also called the visible energy, $E_{ vis}$, is 
given by, 
\begin{equation}
	\label{q1}
	E_{ vis}=\int_0^{T_p} \frac{d E}{1+k_B\langle d E / d x\rangle + C \langle d E / d x\rangle^2}, 
\end{equation}
where $T_p$ is the kinetic energy of the recoiling proton, $\langle d E / d x\rangle$ is the average energy loss 
of proton in the medium, $k_B$ is the Birks' constant \cite{Birks:1951boa}, and $C$ is the bimolecular 
correction \cite{Chao:2021orr}.  The 
proton light-yield of WbLS (formulated from 5\%\,LAB) has been recently measured in Ref. 
\cite{Callaghan:2022ahi} where the quenching parameters have been determined as, 
\begin{align}\label{eq:unc}
	k_B^{\rm WbLS} &= (1.65\pm0.81)\times 10^{-3} \, \text{(cm/MeV)}, \\
	C^{ \rm WbLS} &= (13.30\pm2.70)\times 10^{-6} \, \text{(cm/MeV)}^2. \label{q2}
\end{align}
We approximate $\langle d E / d x\rangle$ as a weighted average of $\rm H_2 O$ and LAB using the 
\rm{PSTAR}\footnote{\url{www.physics.nist.gov/PhysRefData/Star/Text/PSTAR.html}} database. The 
dissolved scintillator concentration in THEIA would be between 0.5\%--5\% depending on the 
performance. In this paper, we only consider the benchmark value of 5\% for which quenching 
measurements are available. For other concentrations, the results can be scaled accordingly. In 
Fig.\,\ref{fig:quenching}, we show $E_{ vis}(T_p)$ for WbLS (i.e, THEIA) along with the 
uncertainty arising from the measured quenching parameters. We also show $E_{ vis}(T_p)$ for 
LAB (i.e, JUNO) based on the measurements of von Krosigk et. al. \cite{vonKrosigk:2013sa} where the 
quenching parameters have been measured with 2-3\% precision. We will assume that the uncertainty 
in quenching parameters of WbLS can also be significantly reduced with future measurements, and 
only use the central values in our Monte Carlo.

\begin{figure}[t]
	\includegraphics[width=0.44\textwidth]{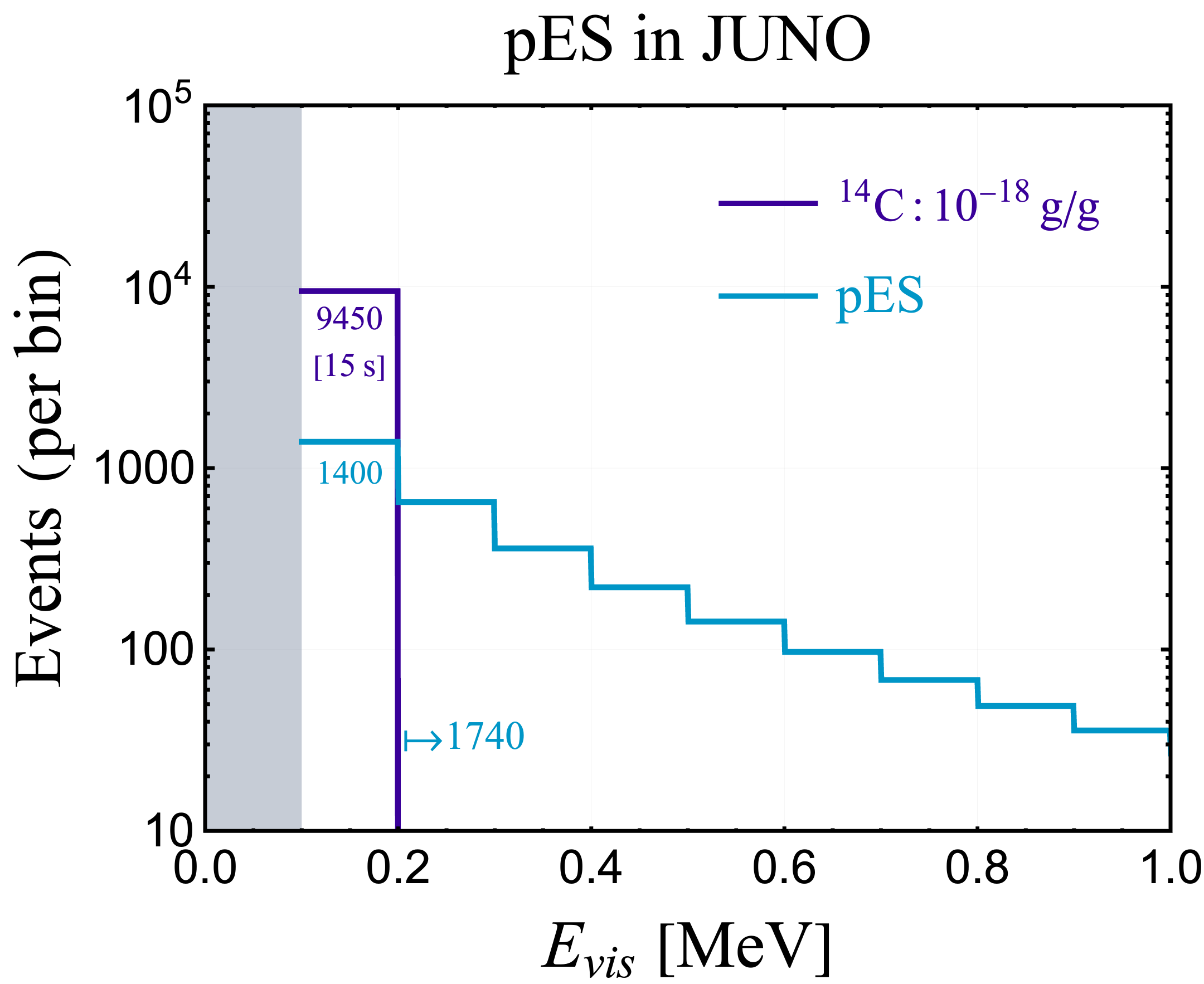}
	\\[0.5cm] 
	\includegraphics[width=0.44\textwidth]{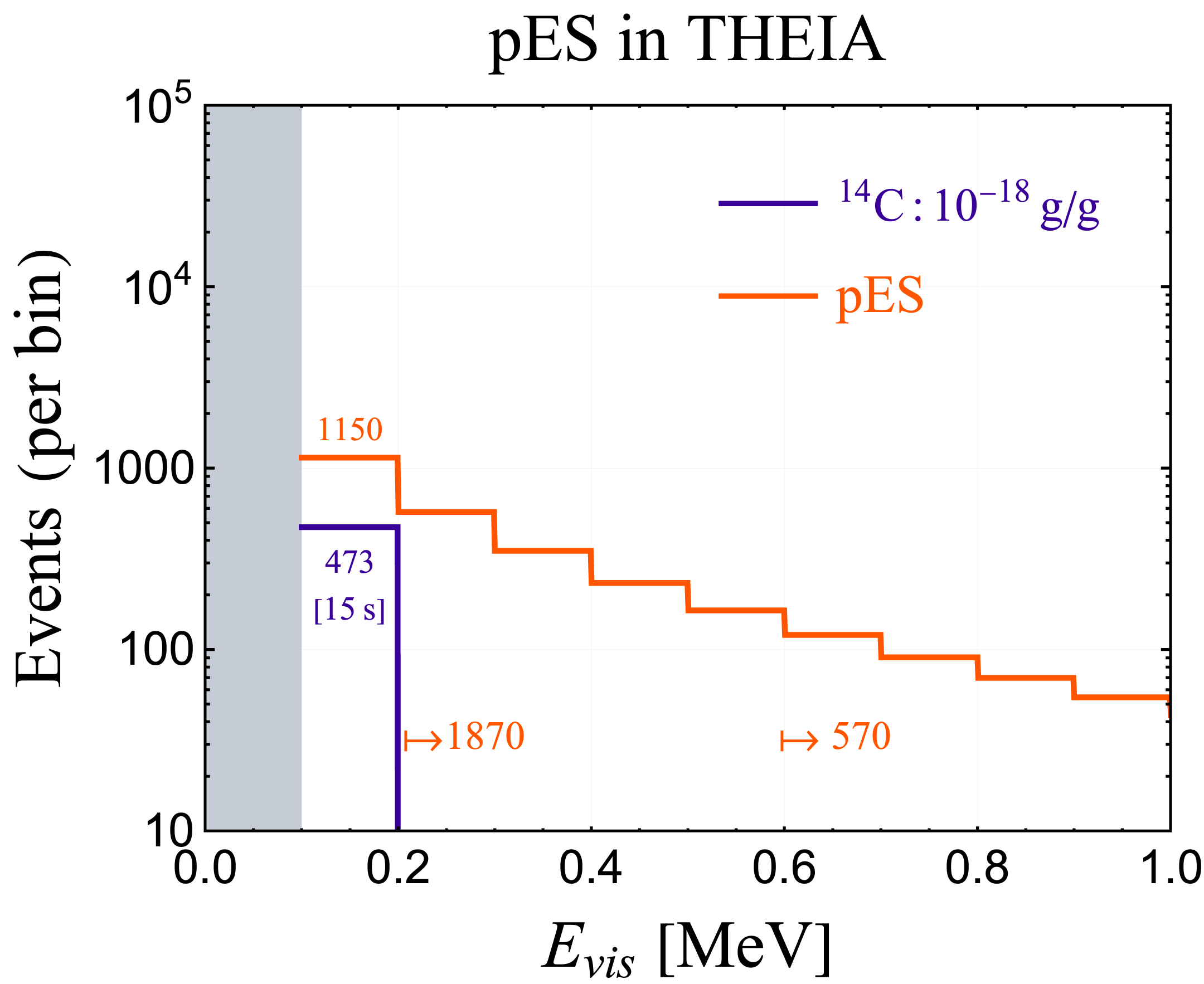}
	\caption{\label{fig:juno_theia} The binned event spectrum for pES from a galactic supernova in JUNO 
	(top) and THEIA (bottom) is shown. The expected number of events from beta-decay of intrinsic 
	$\,^{14}\rm{C}$ over 15\,s interval is also shown for a $\,^{14}\rm{C}$ contamination of $10^{-18}$\,g/g. The 
	gray shaded region represents that $E_{ vis}<0.1$ MeV is dominated by the dark noise 
	of the PMTs. As insets, we have shown the number of events in the first bin, i.e., $E_{ vis} \in$ 
	0.1 -- 0.2 MeV. We also indicate the number of events with $E_{\rm vis}>$\,0.2\,MeV. For THEIA, we 
	also 
	note down the number of events with $E_{ vis}>$\,0.6\,MeV, where it would be possible to 
	distinguish between electrons and protons using Cherenkov to scintillation ratio.}
\end{figure}

It is well understood that the pES event rate is highly sensitive to the $E_{ vis}$ threshold. A natural 
limit for $E_{ vis}$ arises from the dark noise of photo multiplies tubes (PMTs). In JUNO, this 
\emph{dark rate} will overwhelm any signal below 0.1 MeV despite sophisticated trigger schemes 
\cite{Fang:2019lej}. THEIA plans to use a combination of ultrafast PMTs and LAPPDs (Large Area 
Picosecond Photo-Detectors) \cite{Askins:2019oqj, Lyashenko:2019tdj}, which would be different from 
JUNO. Based on 
the trigger schemes and dark noise of individual PMTs and LAPPDs, one can determine the low-energy 
threshold for THEIA. However, this is beyond the scope of this work and we adopt 0.1 MeV as our 
threshold, similar to JUNO.   

Another major background arises from beta-decay of intrinsic $\,^{14}\rm{C}$ \cite{Beacom:2002hs}. 
For JUNO, the estimate for $\,^{14}\rm{C}$ levels is around $10^{-17}$\,g/g, which results in a 
background activity of around 42 kHz/(20 kton)\cite{JUNO:2015zny}.  The scintillation signal of these 
electrons cannot be distinguished from the scintillation of recoiling protons, which results in a wall-like 
background below 0.2 MeV \cite{Beacom:2002hs}. As a result, the threshold for pES events in JUNO is 
often considered to be 0.2 MeV. Ideally, the contamination level of $\,^{14}\rm{C}$ in JUNO would be 
lowered to $10^{-18}$\,g/g, something that is already achieved in Borexino \cite{Borexino:1998eqi}. 
The $\,^{14}\rm{C}$ levels in THEIA is not yet determined. To approximate the $\,^{14}\rm{C}$ 
background in THEIA, we will assume that the scintillator component is extremely radiopure and 
Borexino-like $\,^{14}\rm{C}$ levels can be achieved in WbLS. As there are no long-lived 
radio-isotopes of oxygen, only the scintillator component of WbLS gives rise to large low-energy 
backgrounds. 

Given the fluence, cross section, and quenching parameters, the computation of event rates is 
straightforward. In Fig. \ref{fig:juno_theia}, we show the binned pES event spectrum ($\Delta E_{\rm 
bin} = 0.1$ MeV) for JUNO and THEIA. Our estimates for JUNO agree with the ones reported in Refs. 
\cite{An:2015jdp, Lu:2016ipr, Li:2017dbg, Li:2019qxi}, barring a small enhancement due to weaker 
quenching. Considering smearing due to energy resolution, we estimate that only 15\% of 
$\,^{14}\rm{C}$ beta-decays would have $E_{ vis} \in 0.1$--0.2 MeV, and none above that. Over 
the 15\,s duration of the supernova explosion and for $E_{ vis} \in0.1$--0.2 
MeV, we estimate that JUNO would detect $\sim$9450 events from $\,^{14}\rm{C}$ beta decays, and 
$\sim$1400 events from pES. On the other hand, THEIA would detect only $\sim$475 events from 
$\,^{14}\rm{C}$ beta decays, and $\sim$1150 events from pES. The duration of the supernova burst 
could be $\sim 10$\,s and the $\,^{14}\rm{C}$ backgrounds can be scaled appropriately. The improved 
performance of THEIA can be attributed to the fact that it has a larger proton-to-$\,^{14}\rm{C}$ ratio 
and weaker quenching than JUNO. Lastly, we also note that THEIA would observe $\sim$570 events 
for $E_{ vis} \geq$\,0.6 MeV 
where it would be theoretically possible to distinguish between electron and proton recoils using 
Cherenkov-to-scintillation ratio. However, we do not anticipate any large electron backgrounds at 
these energies.

\begin{table}[t]
	\caption{\label{tab:burst} The expected pES event rates in JUNO and THEIA from the neutronization burst 
	phase of a supernova at 10 kpc are tabulated below. We consider two benchmark progenitor masses, i.e., 
	15 $M_\odot$ and 27 $M_\odot$. The event rates depend on the nuclear equation of state (LS220 
	\cite{Lattimer:1991nc} and 
	Shen \cite{Shen:1998gq}). We show the pES and $\,^{14}\rm{C}$ event rates for $E_{vis} \in 0.1$--0.2 MeV 
	and $E_{vis} > 
	0.2$ MeV. The $\,^{14}\rm{C}$ concentration is optimistically assumed to be $10^{-18}$ g/g. }
	\begin{ruledtabular}
		\begin{tabular}{c c c c c}
			\toprule
			Model & $E_{vis} \left[\rm MeV\right]$ & Event & JUNO & THEIA \\
			\hline  
			\multirow{4}{*}{\shortstack{15 M$_\odot$\\(LS220)}} & \multirow{2}{*}{0.1 -- 0.2} & pES & 9.34 & 8.84 
			\\ 
			& & $\,^{14}\rm{C}$ & 12.60 & 0.63 \\
			\cline{2-5}
			 & \multirow{2}{*}{$>$\,0.2} & pES & 4.71 & 5.96 \\ 
			  & & $\,^{14}\rm{C}$ & -- & -- \\
			 \hline
			\multirow{4}{*}{\shortstack{15 M$_\odot$\\(Shen)}} & \multirow{2}{*}{0.1 -- 0.2} & pES & 9.42 & 8.90 \\ 
			& & $\,^{14}\rm{C}$ & 12.60 & 0.63 \\
			\cline{2-5}
			& \multirow{2}{*}{$>$\,0.2} & pES & 4.98 & 6.24 \\ 
			& & $\,^{14}\rm{C}$ & -- & -- \\		
			\hline
			\multirow{4}{*}{\shortstack{27 M$_\odot$\\(LS220)}} & \multirow{2}{*}{0.1 -- 0.2} & pES & 9.77 & 9.26 
			\\ 
			& & $\,^{14}\rm{C}$ & 12.60 & 0.63 \\
			\cline{2-5}
			& \multirow{2}{*}{$>$\,0.2} & pES & 4.95 & 6.26 \\ 
			& & $\,^{14}\rm{C}$ & -- & -- \\			
			\hline
			\multirow{4}{*}{\shortstack{27 M$_\odot$\\(Shen)}} & \multirow{2}{*}{0.1 -- 0.2} & pES & 9.98 & 9.40 
			\\ 
			& & $\,^{14}\rm{C}$ & 12.60 & 0.63 \\
			\cline{2-5}
			& \multirow{2}{*}{$>$\,0.2} & pES & 5.31 & 6.65 \\ 
			& & $\,^{14}\rm{C}$ & -- & -- \\	  
			\bottomrule
		\end{tabular}
	\end{ruledtabular}
	
\end{table}

The neutrinos from a core-collapse supernova can be temporally separated in three phases: the 
\emph{neutronization burst, accretion}, and \emph{cooling}. The neutronization burst phase is the first 
$\sim$20\,ms interval where a prompt burst of $\nu_e$ occurs due to electron capture by free protons in the 
stellar nucleus. Unlike other phases, the neutronization burst phase has a very weak dependence on the 
progenitor properties, and is almost a \emph{standard candle}. It has been proposed that the neutronization 
burst can be used to determine the supernova distance \cite{Kachelriess:2004ds,Segerlund:2021dfz}, 
absolute neutrino mass 
\cite{Zatsepin:1968kt, Nardi:2003pr, 
Nardi:2004zg,Pagliaroli:2010ik,Lu:2014zma,Rossi-Torres:2015rla,Pompa:2022cxc}, neutrino mass ordering 
\cite{Brdar:2022vfr}, as well as new physics scenarios \cite{Huang:2021enl, Tang:2020pkp}. In Tab. 
\ref{tab:burst}, we provide our estimated event 
rates from pES in JUNO and THEIA for a 
supernova at 10 kpc. We use the time-dependent flux parameters provided in EstrellaNueva 
\cite{Gonzalez-Reina:2022ehy} which relies on 
the simulations of Garching group \footnote{\url{https://wwwmpa.mpa-garching.mpg.de/ccsnarchive/}} 
\cite{Mirizzi:2015eza}. We 
choose two benchmark progenitor masses, i.e., 15 $M_\odot$ and 27 $M_\odot$ and provide estimates using 
two equation of state, i.e., LS220 \cite{Lattimer:1991nc} and Shen \cite{Shen:1998gq}. We find that for 
$E_{vis}>0.2$\,MeV, where there are no $\,^{14}\rm{C}$ backgrounds, we get $\sim5$ pES events in JUNO 
and $\sim6$ pES events in THEIA. We 
also note that for $E_{vis}\in0.1$--0.2\,MeV, we get $\sim9$ pES events in JUNO and THEIA, and the 
$\,^{14}\rm{C}$ backgrounds\footnote{assuming $\,^{14}\rm{C}$ concentration of $10^{-18}$ g/g} are 
$\sim12$ and $\sim 5$ in JUNO and THEIA respectively. A key message from this estimate is that for the first 
$\sim 20$ ms of a supernova, the pES event rates in THEIA will overwhelm the $\,^{14}\rm{C}$ backgrounds, 
whereas for JUNO, the event rates are comparable. This would have interesting applications for supernova 
early warning system (SNEWS) \cite{SNEWS:2020tbu} where pES events may be used as secondary trigger. 
 
\section{Extracting $\Delta s$ from $\rm{pES}$}\label{sec:extract}

In order to reconstruct $\Delta s$ from pES events in JUNO and THEIA, one needs to independently 
determine the flux of all flavors of supernova neutrinos. As mentioned earlier, we will rely on other 
neutrino detectors to provide information on fluence parameters using interactions that do not depend 
on $\Delta s$. It has been 
demonstrated in Ref. \cite{Nikrant:2017nya} that DUNE can measure $\nu_e$ fluence parameters 
through charged-current interactions with argon. On the other hand, $\bar{\nu}_e$ and $\nu_x$ 
parameters can be determined using water Cherenkov detectors through inverse beta decay, elastic 
scattering with electrons, and interactions with oxygen \cite{GalloRosso:2017mdz}. It must be noted 
that, Gadolinium loaded Hyper-Kamiokande can attain sensitivity to $\nu_e$ similar to DUNE, and all 
fluence parameters can be determined from a single experiment \cite{Laha:2013hva}. Similarly, in Ref. 
\cite{Laha:2014yua} it is proposed to detect $\nu_e$ using interactions on carbon in scintillator detectors. 
However, both methods rely on statistical subtraction of $\nu_x$ events that would be measured using pES in 
JUNO-like detectors, and hence indirectly depend on $\Delta s$. Thus, we argue that DUNE is essential for 
this analysis, and assume that it would be operational before the next galactic supernova. We 
consider both possibilities where Super-Kamiokande (SK) may or may-not be upgraded to 
Hyper-Kamiokande (HK). The uncertainties in reconstruction of fluence parameters using these detectors is 
summarized in Tab. \ref{tab:errors}. We use the DUNE sensitivity to $\nu_e$ from Ref. \cite{Nikrant:2017nya} 
and SK/HK sensitivity to $\bar{\nu}_e$ and $\nu_x$ from Ref. \cite{GalloRosso:2017mdz}. 

\begin{table}[b]
	\caption{\label{tab:errors} The fractional symmetric uncertainties ($1\sigma$) on the flux parameters 
	for a galactic supernova at 10 kpc are tabulated below. The uncertainties in $\nu_e$ parameters 
	measured with DUNE are 
	adopted from Ref. \cite{Nikrant:2017nya}. The uncertainties in $\bar{\nu}_e $ and $\nu_x$ parameters
	measured by Super-Kamiokande (SK) or Hyper-Kamiokande (HK) are taken from Ref. 
	\cite{GalloRosso:2017mdz}. The second column gives the central value of these parameters.}
\begin{ruledtabular}
		\begin{tabular}{c c c c}
			\toprule
			Parameter & Cen. & DUNE+SK & DUNE+HK \\
			\hline
			$\mathcal{E}_{\nu_e}$ [$10^{53}$ erg]& 0.5 & 13\% &  13\% \\
			$\mathcal{E}_{\bar{\nu}_e}$ [$10^{53}$ erg]& 0.5 & 10\% &  4\% \\
			$\mathcal{E}_{\nu_x}$ [$10^{53}$ erg]& 0.5 & 18\% &  8\% \\
			\hline
			$\langle E_{\nu_e} \rangle$ [MeV]& 12 & 7\% &  7\% \\
			$\langle E_{\bar{\nu}_e} \rangle$ [MeV] & 14 & 6\% &  3\% \\
			$\langle E_{\nu_x} \rangle$ [MeV] & 16 & 11\% &  7\% \\
			\hline
			$\alpha_{\nu_e}$ & 3.0 & 18\% &  18\% \\
			$\alpha_{\bar{\nu}_e}$ & 3.0 & 20\% &  7\% \\
			$\alpha_{\nu_x}$ & 3.0 & 21\% &  17\% \\
			\bottomrule
		\end{tabular}
	\end{ruledtabular}

\end{table}

\begin{figure}[b]
	\includegraphics[width=0.4\textwidth]{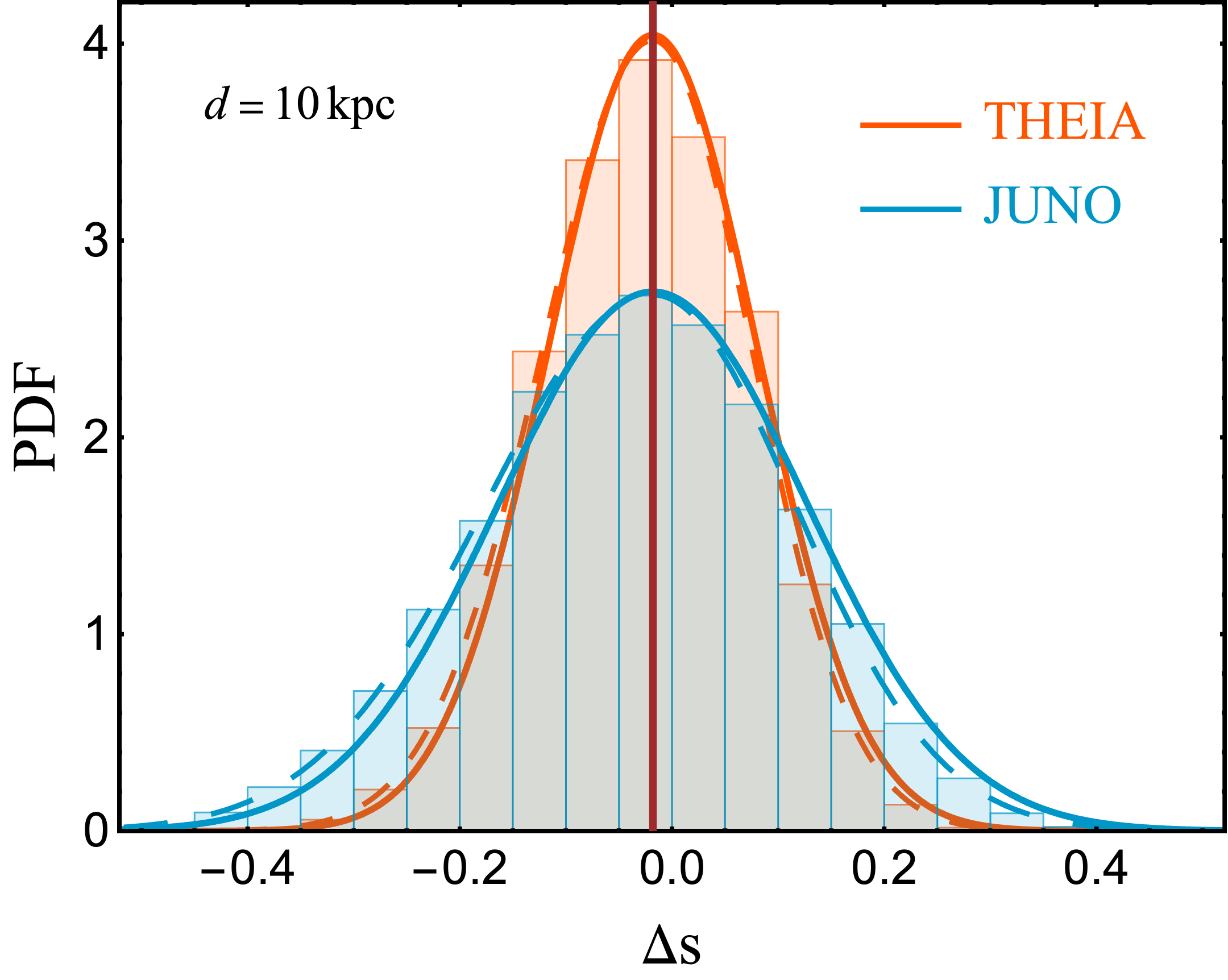}
	\caption{\label{fig:asym} The histogram of reconstructed $\Delta s$ in one of the MC simulations
	JUNO (blue bars) and THEIA (orange bars) using uncertainties from DUNE+HK scenario and for a supernova 
	at 10 kpc is shown. The vertical maroon line denotes the assumed true-value of $\Delta s$ in the simulation 
	taken from Ref. \cite{Chambers:2015bka}. The overlaid solid curve is a Gaussian fit with symmetric 
	errors, and dashed curve is a fit with asymmetric errors. As the difference is very small, we only report 
	the symmetric uncertainty in this paper. }
\end{figure}

We use the technique of uncertainty quantification using Monte Carlo simulations to estimate the sensitivity of 
pES in JUNO and THEIA to $\Delta s$. Although we can, in principle, look at the spectrum of events, instead 
we only use the total number of pES events to estimate $\Delta s$. For JUNO, we use the $E_{vis}$ threshold 
at 0.2\,MeV, and for THEIA we use $E_{vis}$ threshold at 0.1\,MeV. We generate a collection of 
$10\times10^4$ sample observations for the four scenarios. For each sample, the fluence parameters are 
randomly chosen 
from a Gaussian distribution with mean and variance from Tab. \ref{tab:errors}. The true value of $\Delta s$ in 
the simulation is chosen to be $0.018$, which is the central value in Ref. \cite{Chambers:2015bka}. 
This choice has negligible impact on our results. The 
supernova distance is strongly correlated with the total neutrino luminosity. Since we are using the 
reconstructed parameters, one can consider that the uncertainty in $\mathcal{E}_\nu$ in Tab. 
\ref{tab:errors} is in fact the uncertainty in $\mathcal{E}_\nu/d^2$. The distribution of estimated 
number of events in each scenario is asymmetric with variance that is consistent with our expectation of a 
Gaussian 
distribution. For each sample, we determine the expected $\Delta s$ using a simple fit. The 
collection of $\Delta s$ thus obtained is used to determine the uncertainty which is reported in Tab. 
\ref{tab:result}. Note that the distribution of $\Delta s$ obtained this way is also asymmetric, but the 
asymmetry is small and can be ignored. This can be clearly seen in Fig. \ref{fig:asym} 
where we show the result of one of the realizations. 

\begin{table}[t]
	
	\caption{\label{tab:result} The symmetrized $1\sigma$ uncertainties on the $\Delta s$ obtained 
		from a fiducial galactic supernova are tabulated for three choices of distance. The $\nu_e$ spectrum is 
		determined from DUNE. The two columns refer to the two scenario where $\bar{\nu}_e$ and $\nu_x$ are 
		determined from gadolinium doped Super-Kamiokande (SK) or Hyper-Kamiokande (HK). The pES events 
		can 
		be measured using either JUNO or THEIA. The higher sensitivity of THEIA is due to weaker quenching and 
		lower backgrounds than JUNO.}
	\begin{ruledtabular}
		\begin{tabular}{ c c c c}
			\toprule
			Distance & Detector & DUNE+SK & DUNE+HK \\
			\hline
			\multirow{2}{*}{10 kpc} & JUNO & $\pm 0.223$ & $\pm 0.146$\\
			&THEIA & $\pm 0.160$ & $\pm 0.099$ \\
			\hline
			\multirow{2}{*}{5 kpc} & JUNO & $\pm 0.117$ & $\pm 0.078$\\
			&THEIA & $\pm 0.085$& $\pm 0.053$ \\
			\hline
			\multirow{2}{*}{1 kpc} & JUNO & $\pm 0.024$ & $\pm 0.015$\\
			&THEIA & $\pm 0.017$& $\pm 0.010$ \\
			\bottomrule
		\end{tabular}
	\end{ruledtabular}
\end{table}

For a supernova at 10 kpc, the reconstructed spectra of neutrinos has large uncertainty and $\delta \Delta s$ 
is around $\pm 0.2$. These limits are slightly better than the beam neutrino experiments, but weaker than 
pDIS experiments. An interesting possibility, where the uncertainties in reconstructed fluence parameters 
would be smaller, is to consider a nearby supernova. It is estimated that there is a $\sim$10\% chance that the 
next galactic supernova is within 5 kpc of the earth and $\sim$0.25\% chance that is within 1 kpc 
\cite{Adams:2013ana}. For more promising estimates, one can refer to Refs \cite{Schmidt, The:2006iu}. 
Notably, there are $\sim$30 core-collapse supernova candidates within 1 kpc \cite{Mukhopadhyay:2020ubs}. 
To approximate the reconstruction efficiency of DUNE and SK/HK for a nearby supernova, we will assume that 
the systematic uncertainty (mainly arising from cross-sections) is small and the errors mentioned in Tab. 
\ref{tab:errors} are mostly statistical. For a supernova at a distance $d$, we scale the uncertainties by a factor 
of $N^{-1/2} \sim d$ to obtain our estimates. One can consider our projected sensitivity for these supernovae 
to be optimistic. A careful re-analysis of reconstruction for nearby supernova is required, but beyond the 
scope of this work. The $\delta \Delta s$ obtained have been tabulated in Tab. \ref{tab:result} for three 
benchmark distances of 10, 5, and 1 kpc. We present our results for two scenarios, with and without 
Hyper-Kamiokande. The uncertainties reported in Tab. \ref{tab:result} for various cases are shown in Fig. 
\ref{fig:main} along with results from other experiments for comparison.  

In the method described above, one is limited by the ability to properly reconstruct the fluence parameters 
using other detectors. A promising possibility is to exploit the 
neutronization burst phase of the supernova. The neutrino flux from the neutronization burst is very well 
predicted through simulations, and does not depend heavily on the properties of the progenitor. However, we 
find that there are three challenges in reconstructing $\Delta s$ using neutronization burst. First, despite being 
a \emph{standard candle}, there are large (5-10\%) systematic uncertainties in the flux estimates. The 
differences arise partly from various nuclear physics parameters, and partly from the unknown mass of the 
progenitor \cite{Kachelriess:2004ds}. This is the source of differences in event rates given in Tab. 
\ref{tab:burst}. Second, if the supernova is obscured by galactic dust, the distance to the 
supernova will be poorly known. As $\Delta s$ and $d$ both have similar effects on pES event rates, they are 
strongly correlated and the reconstructed $\Delta s$ would have large uncertainties as well. Lastly, only a 
small fraction of the total luminosity is emitted during the neutronization burst. The pES event rates in 
JUNO and THEIA are small (cf. Tab. \ref{tab:burst}), and result in large statistical uncertainty. Our preliminary 
estimates suggest that the reconstructed $\Delta s$ is not significantly better than the one obtained by the 
aggregate pES events for a supernova at 10 kpc. For a nearby supernova, the uncertainty in the distance 
would be much smaller. A dedicated study of reconstruction of spectra of neutrinos from neutronization burst 
in DUNE and SK/HK would be required for a more refined estimate the $\Delta s$-sensitivity. We look forward 
to future work in this direction. 

\section{Summary}\label{sec:summary}

A core collapse supernova in our galaxy would provide an excellent opportunity to study low-energy 
physics. In this paper, we looked at the possibility whether the strange spin of proton, $\Delta s$, could 
be measured using neutrinos from the next galactic supernova, and the results look promising. We 
utilize the ability of near-future detectors DUNE and Hyper-Kamiokande to reliably estimate the 
spectra of \emph{all} flavors of supernova neutrinos using interactions that do not depend on $\Delta 
s$. The reconstructed spectra can be used to estimate the pES event rate in large-volume scintillation 
detectors such as JUNO and/or THEIA, and compared with the measured event rates. Assuming that 
the variation only arises because of an unknown $\Delta s$, it is straightforward to reconstruct the 
allowed values of $\Delta s$. 

To estimate the sensitivity of JUNO and THEIA, we perform a simple Monte Carlo simulation for 
uncertainty quantification.  For a supernova at 10 kpc, we estimate that JUNO would be able to  
constrain $\Delta s$ within $\pm 0.15$ ($\pm 0.22$) using the reconstructed fluence parameters from 
DUNE and Hyper-(Super-)Kamiokande;  whereas THEIA would be able to constrain $\Delta s$ within 
$\pm 0.10$ ($\pm 0.16$) using DUNE and Hyper-(Super-)Kamiokande. The better performance of 
THEIA is due to larger light yield of water-based liquid scintillator as compared to linear alkyl benzenes, and 
relatively larger proton-to-$\,^{14}\rm{C}$ ratio that allows for a lower threshold. We also provide estimates 
for the neutronization burst phase, and note that for the brief $\sim$20 ms window, the pES event rates in 
THEIA would be larger than the $\,^{14}\rm{C}$ background. Moreover, the pES event rates from 
neutronization burst in JUNO are comparable to $\,^{14}\rm{C}$ background. We propose that pES events 
could also be used in supernova early warning systems as secondary or fail-safe triggers.

If we are lucky and the next galactic supernova is within 1 kpc, then the projected sensitivity improves to 
approximately $\pm 0.01$ which is comparable to results from lattice QCD, and better than polarized 
deep-inelastic scattering experiments. Such a nearby supernova would provide us with a 
once-in-a-lifetime opportunity to measure the true $Q^2\rightarrow0$ limit of neutrino-proton scattering, 
without a dedicated experiment. The rare circumstance and the reliance on multiple large-volume detectors is 
an attestation to the difficulty of experimental determination of $\Delta s$. \\

\section*{Acknowledgments}
The author would like to thank Ranjan Laha, Basudeb Dasgupta, Mary Hall Reno, Vedran Brdar, and Xunjie Xu 
for their useful comments on the manuscript. This work is supported in part by US Department of Energy 
grant DE-SC-0010113. The author also acknowledges the support of TIFR (India) where a part of this work 
was completed. 

\bibliography{SN_probe_references.bib}

\end{document}